\documentclass[11pt, a4paper]{article}
\usepackage[margin=2.5cm]{geometry}
\usepackage{authblk}
\usepackage[utf8]{inputenc}
\usepackage{graphicx}
\usepackage{subcaption}
\usepackage[section]{placeins}
\usepackage{amsmath}
\newcommand{\bmu}{\mbox{\boldmath $\mu$}}
\DeclareMathOperator*{\argmax}{arg\,max}

\usepackage{xcolor}

\usepackage{cancel} 

\title{Source location on multilayer networks}
\author[1]{R. Paluch}
\author[1]{Łukasz G. Gajewski}
\author[1]{K. Suchecki}
\author[1,2]{J. A. Hołyst}

\affil[1]{Faculty of Physics, Warsaw University of Technology, Koszykowa 75, 00-662 Warszawa, Poland}
\affil[2]{ITMO University, Kronverkskiy Prospekt 49, St Petersburg, Russia 197101}

\begin{document}

\maketitle

\begin{abstract}
Nowadays it is not uncommon to have to deal with dissemination on multi-layered networks and often finding the source of said propagation can be a crucial task.
In this paper we tackle this exact problem with a maximum likelihood approach that we extend to be operational on multi-layered graphs.
We test our method for source location estimation on synthetic networks and outline its potential strengths and limitations.
We also observe some non-trivial and perhaps surprising phenomena where the more of the system one observes the worse the results become whereas increased problem complexity in the form of more layers can actually improve our performance.
\end{abstract}

\section{Introduction}
Sharing and spreading information is one of the cornerstones of civilization.
Not all information is valuable however, and some, such as misinformation or conspiracy theories can have detrimental effect on the society as a whole.
In practice the question of the true source of such is very important, because it allows for developing or implementing appropriate preventive measures.

The question of how to find the source of information or a rumour is not new and have seen large amount of research devoted to this topic.
From methods based on single time snapshot \cite{Shah2011, Prakash2012, Lokhov2014, Zhu2016b} to observer-based \cite{pintolocating, Karamchandani2013, Luo2014} where we know information from only limited set of nodes, but it spans whole or at least part of the time of the whole spreading process.
More recent works aim at relaxing the assumptions about the process and its parameters \cite{wanglocating, sheanovel} to allow finding a source of a signal without prior knowledge of the actual spreading process.
There have been also works considering source finding problem on a variable, temporal topology of connections \cite{huanglocating, jiangrumor}.
The multi-layer nature of human interactions in online and offline environments has not been considered in current research on the topic.

Multiplex of several layers \cite{boccalettithestructure, kivelamultilayer} are usually either typical multiplex networks, with nodes connected by edges of various natures, and multi-layer networks, where the nodes themselves are split into the layers and may have different states or properties in each layer or sometimes not exist in some at all.
In this paper we have considered the latter type, assuming that online users, while being the same people can exhibit different preferences, opinions and behaviours in different social networks they belong to.
It follows that they may not relay information learned in one network to another or do so with a delay.
Thus, users are present in several layers, in each having its own independent state with regard to knowing information, but they are connected by self infer-layer links.
We consider the situation where there exists a number of observers - users that report the time at which they received spreading message, information or a rumour.
We explore the question of locating a single, true source of the rumour in a multi-layer network, knowing the connection topology and times at which the message arrived at select set of observer nodes.

\section{Preliminaries}

\subsection{Multilayer graphs}
\begin{figure}[!hbt]
    \centering
    \begin{subfigure}[b]{0.48\textwidth}
        \caption{$t=0$}
        \includegraphics[width=\textwidth]{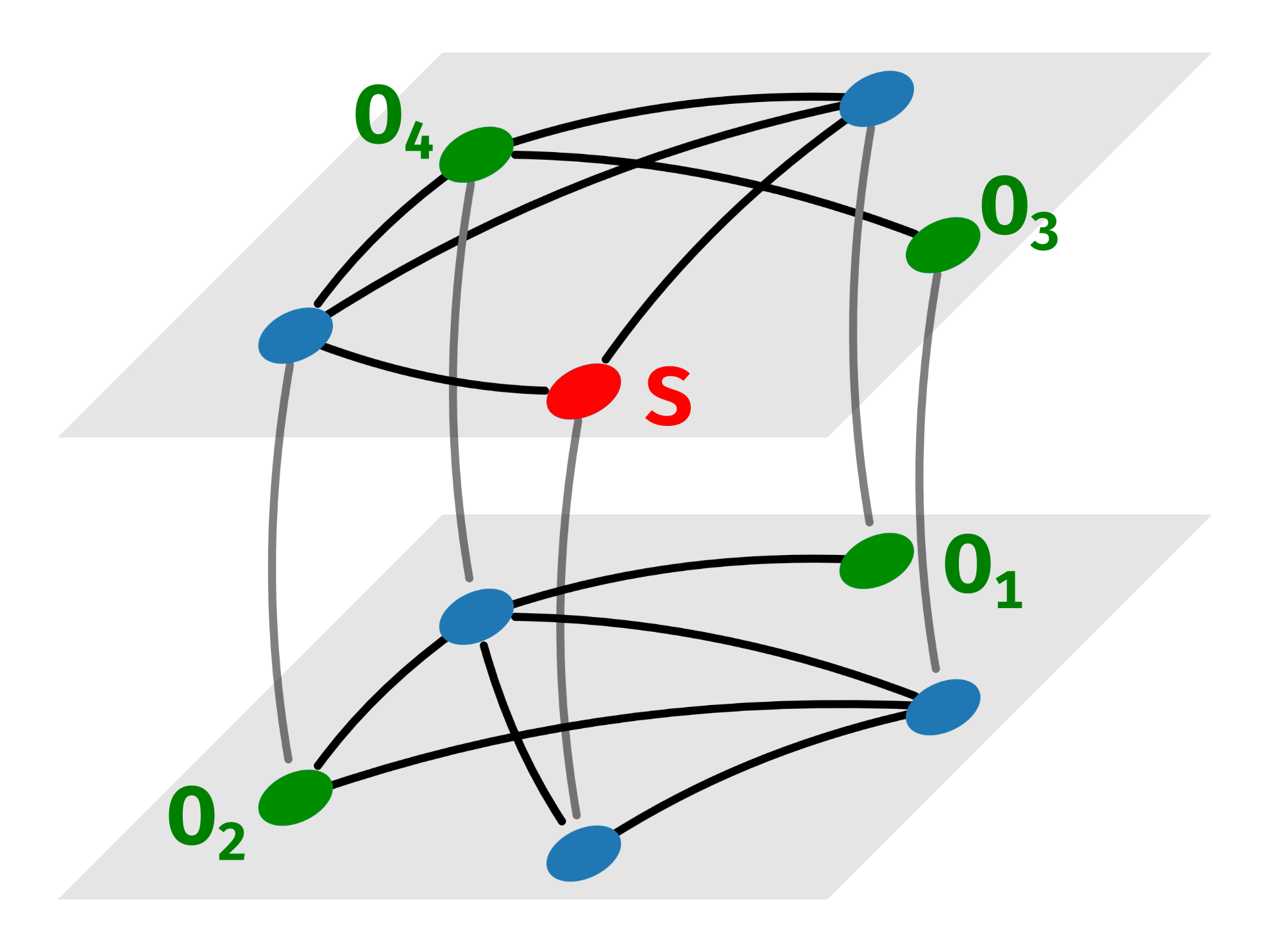}
    \end{subfigure}
     \hfill
    \begin{subfigure}[b]{0.48\textwidth}
        \caption{$t>0$}
        \includegraphics[width=\textwidth]{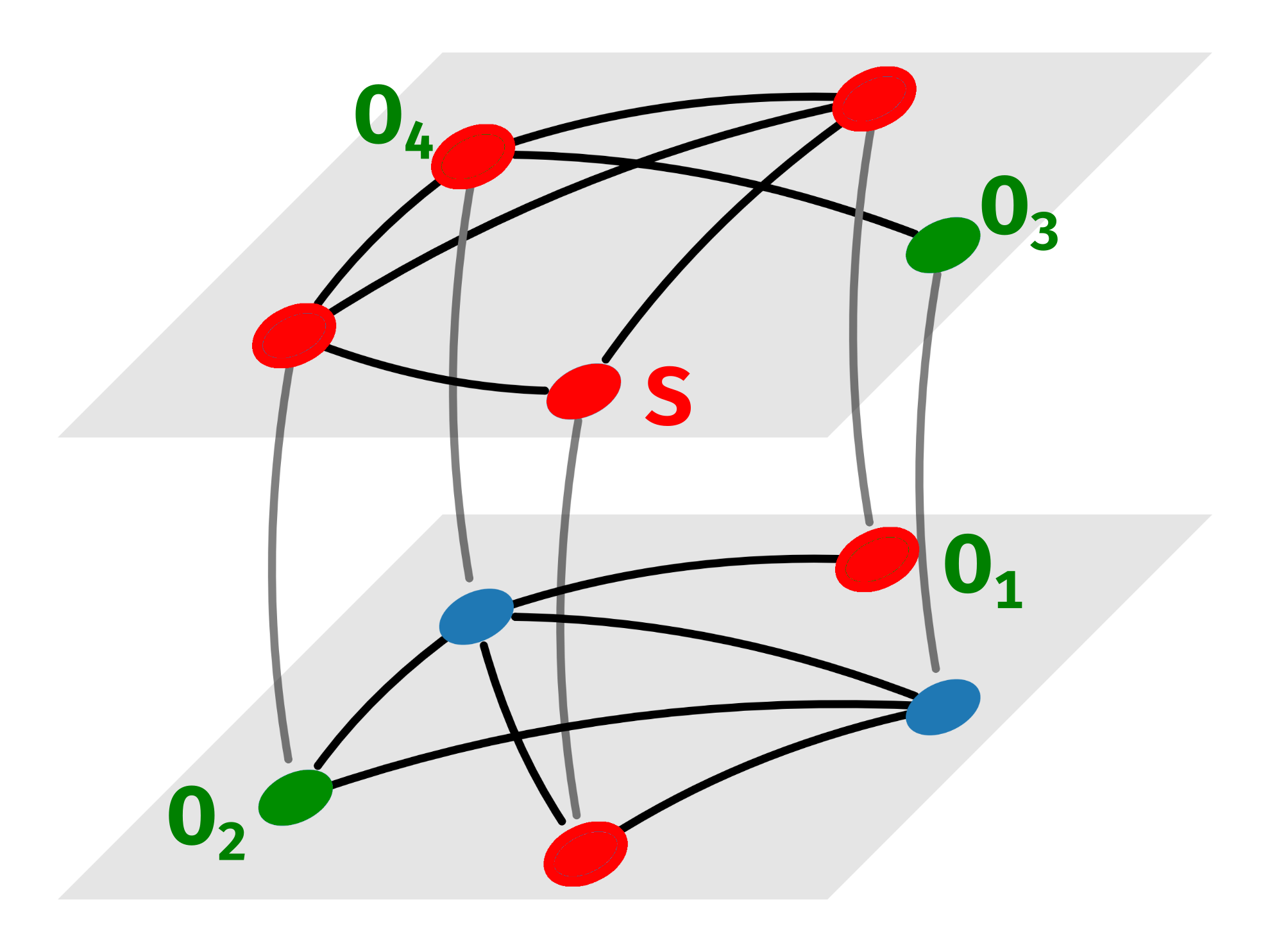}
    \end{subfigure}
    \caption{Schematic illustration of studied systems. Two snapshot of a propagation process are presented one in the very beginning $t=0$ (left) and the other at some later time $t>0$ (right). The red nodes are infected with the one labelled $S$ being the source. Blue and green nodes are susceptible where green ones (and labelled with $O_i$) are the observers. This example shows the idea of the setting used in our experiments - a multiplex structure yet the states are not shared between images of nodes. E.g. Both $O_1$ and its image are infected in the right panel, however, while $O_4$ has its replica has not received the signal yet. Additionally it is worth noting that an observer in one layer is not necessarily an observer in the other.}
    \label{fig:multiplex_example}
\end{figure}
In this paper we consider a multi-layer graph with $L$ denoting number of layers. Each layer has $n_l$ nodes and $m_i$ edges where $i \in [1, 2\dots L]$.
The total number of nodes in the network is $n_{tot}=L n_l$.
A given layer's topology is independent from other layers, however, can be potentially correlated (more on that later).
We use the so called \textit{multiplex} topology scheme, i.e. every node is connected to all its replicas (images) in other layers and no other inter-layer links are possible, however, we do not assume that the \textit{states} are shared among layers - given node can have different state than its images. 
See Fig.~\ref{fig:multiplex_example} for an example.

We conduct our studies on two well-known synthetic network models: Erd\H{o}s–R\'{e}nyi (ER) and Barab\'{a}si-Albert (BA). As mentioned before each layer is independent, i.e. we construct $L$ realizations of a given graph model and couple them accordingly with inter-layer links. While every layer is different from others the degree distributions in the BA model are highly correlated - a hub in one layer is most likely a hub in another.

\subsection{Source location}
\begin{figure}[!hbt]
    \centering
    \includegraphics[width=\textwidth]{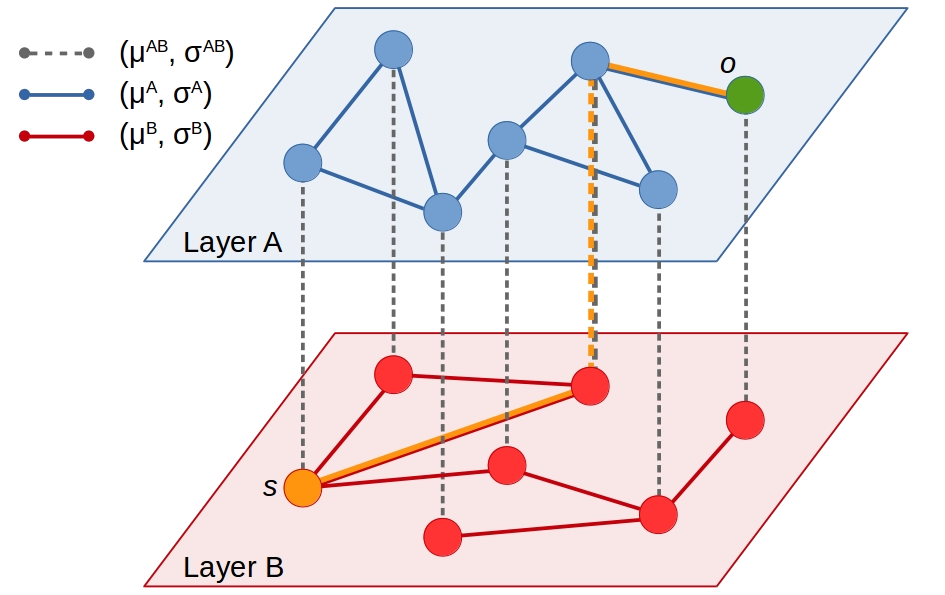}
    \caption{Schematic illustration of the information traversal dynamics. The orange node labelled $S$ is the source of the spread and the green one (labelled $O$) is an observer. Each layer has its own set of propagation properties (mean - $\mu$ - and variance - $\sigma$ - of the traversal time) and the inter-layer coupling is also independent with it respective parameters. The orange colour on edges indicates a shortest path from the source to the observer.}
    \label{fig:duplex_propagation}
\end{figure}
We estimate the source location using detectors-based maximum likelihood estimator.
This method was first introduced by Pinto et al. \cite{pintolocating} for single-layer graphs.
In this work we extend this algorithm to be able to locate the source in the multi-layer structures with different diffusion properties for each layer.
We assume that the distributions of delays on the links in each layer, which describe a spreading process, have finite means $\boldsymbol{\mu} = [\mu_1, \mu_2, \dots, \mu_L, \mu_{inter}]$ and variances $\boldsymbol{\sigma^2} = [\sigma^2_1, \sigma^2_2, \dots, \sigma^2_L, \sigma^2_{inter}]$, where $\mu_{inter}$ and $\sigma^2_{inter}$ describe the distribution of delays on all interlinks between layers.
Moreover, the method requires that all these means and variances are known, as well as all links in each layer.
Also, we have the ability to monitor the states of some preselected replicas $o_i \in O$, called observers.
Please note, that observing a replica in some layer does not mean that we monitor the states of corresponding replicas in the others layers.
One observer is assigned to one replica in one layer.
From infection times reported by observers we construct an observed delay vector $\mathbf{d}$:
\begin{equation}
    \mathbf{d} = (t_2 - t_1, t_3 - t_1, \dots, t_{b} - t_1)^T
\end{equation}
where $b$ is the number of observers (budget), $t_i$ is an infection time of observer $o_i \in O$, and $t_1$ is the infection time of a \textit{reference observer} $o_1$ that is needed here since the $t_0$ is unknown.

\begin{figure}[!hbt]
    \centering
    \includegraphics[width=\textwidth]{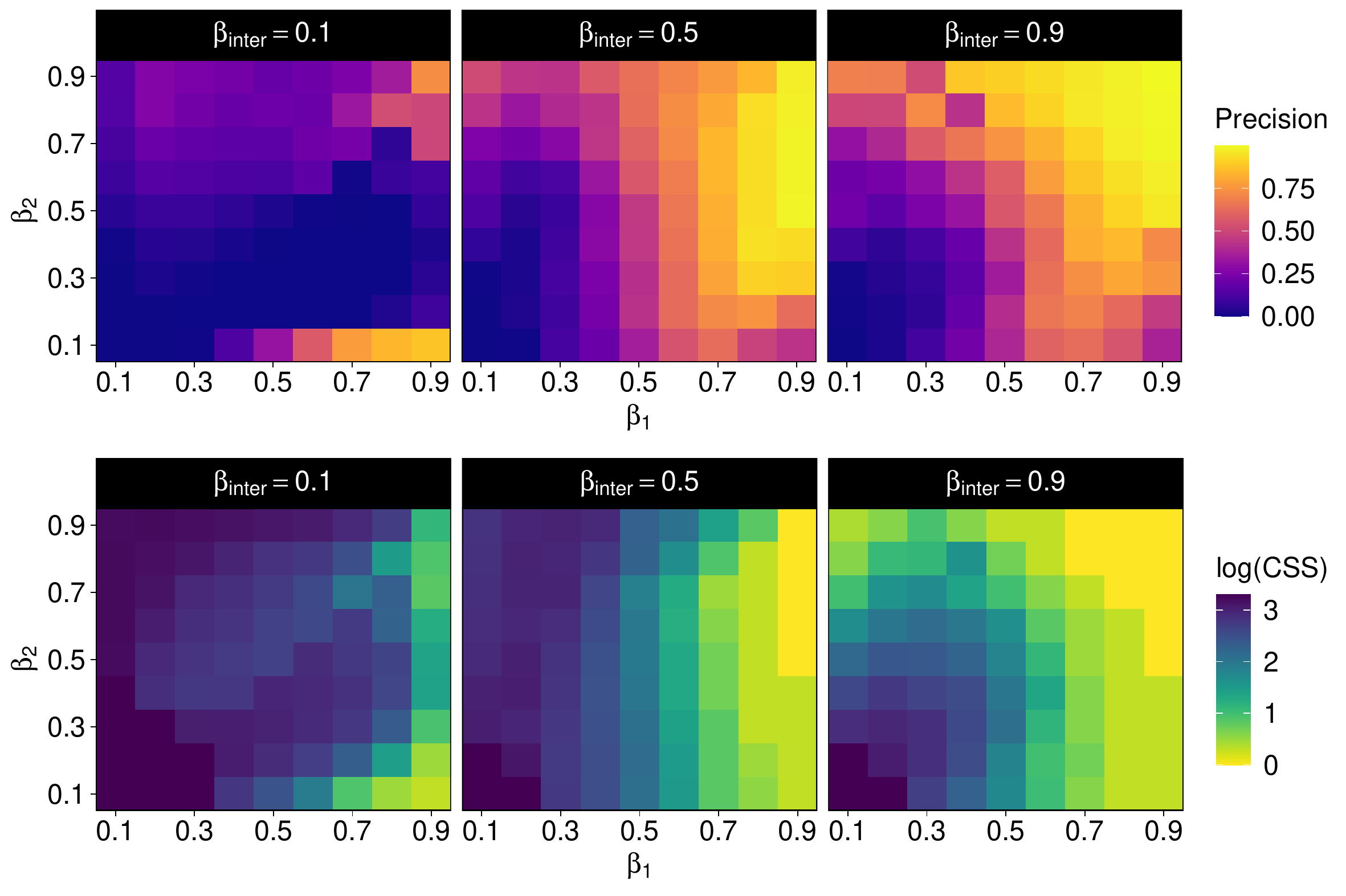}
    \caption{Average precision (higher is better) and 0.95-CSS (lower is better) of source localization in Erd\H{o}s–R\'{e}nyi graph with $L=2$, $n_1=n_2=1000$ and $\langle k_1 \rangle = \langle k_2 \rangle = 8$.
    The observers are placed randomly with equal density in both layers $\rho_1 = \rho_2 = 0.1$.
    Layer 1 with spreading rate $\beta_1$ is a source layer.
    We consider three values of interlayer spreading rate $\beta_{inter}$: 0.1 (left), 0.5 (centre) and 0.9 (right).
    The evaluation metrics are computed from $10^3$ realizations.}
    \label{fig:er_rate_heatmaps}
\end{figure}

\begin{figure}[!hbt]
    \centering
    \includegraphics[width=\textwidth]{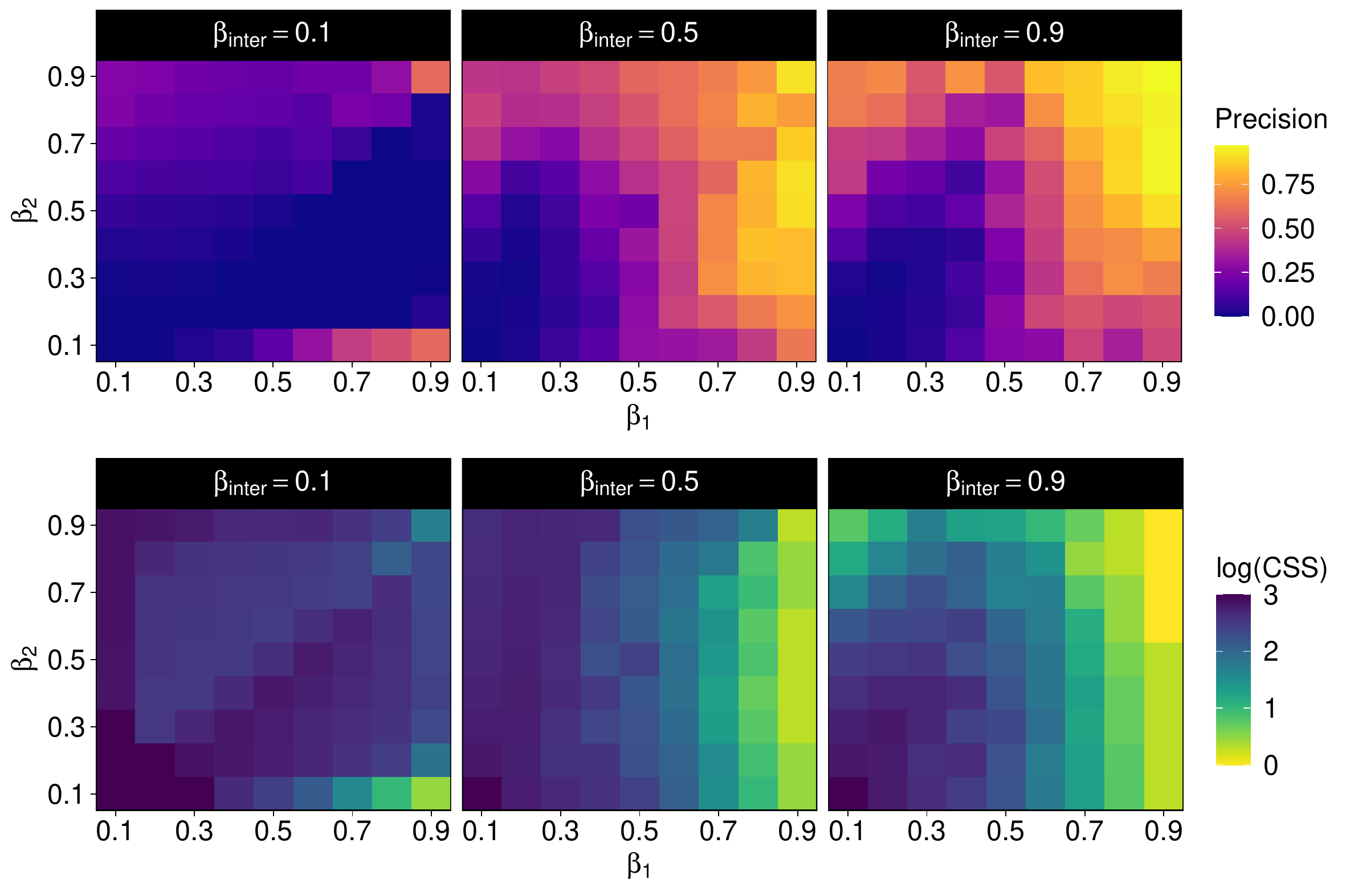}
    \caption{Average precision (higher is better) and 0.95-CSS (lower is better) of source localization in Barab\'{a}si-Albert graph with $L=2$, $n_1=n_2=500$ and $\langle k_1 \rangle = \langle k_2 \rangle = 8$.
    The observers are placed randomly with equal density in both layers $\rho_1 = \rho_2 = 0.1$.
    Layer 1 with spreading rate $\beta_1$ is a source layer.
    We consider three values of interlayer spreading rate $\beta_{inter}$: 0.1 (left), 0.5 (centre) and 0.9 (right).
    The evaluation metrics are computed from $10^3$ realizations.}
    \label{fig:ba_rate_heatmaps}
\end{figure}

Then, for each node $v \in V$ we construct a tree $\mathcal{T}_v$ from the shortest weighted paths (which may contain inter and intra links) between $v$ and all observers $o_i \in O$.
The weights of the links depend on the layer and are given by the vector $\boldsymbol{\mu} = [\mu_1, \mu_2, \dots, \mu_l, \mu_{inter}]$, see Fig.~\ref{fig:duplex_propagation} for illustration.
The rest of the following computations for node $v$ is performed on tree $\mathcal{T}_v$, not on the general graph $\mathcal{G}$.

To obtain the likelihood for node $v$ of being a source we need to compute a deterministic delay vector $\bmu_v$ and the covariance matrix $\mathbf{\Lambda}_v$:
\begin{equation}
\label{eq:det_delay}
    \bmu_v[\,i\,] = |\mathcal{P}(v, o_{i+1})|_{\mu} - |\mathcal{P}(v, o_1)|_{\mu} \qquad i = 1,2,\dots b-1,
\end{equation}
\begin{equation}
    \mathbf{\Lambda}_{v}[\,i,j\,] = |\mathcal{P}(o_{i+1}, o_1) \cap \mathcal{P}(o_{j+1}, o_1)|_{\sigma^2} \qquad i,j = 1,2,\dots b-1,
\end{equation}
where $\mathcal{P}(v, o_i)$ denotes the path (a set of links) between nodes $v$ and $o_i$ in the tree $\mathcal{T}_v$, while $A \cap B$ means a set of shared links between paths A and B.
The operators $|\,\mathcal{P}\,]_{\mu}$ and $|\,\mathcal{P}\,]_{\sigma^2}$ denotes respectively summing up the mean delays or the variances of delays on links in the path $\mathcal{P}$.

Finally we compute a \textit{score} for each node $v$ and use maximum likelihood rule to determine the most probable source of the epidemic $\hat{s}$:
\begin{equation}
\label{eq:argmax}
    \hat{s} = \argmax_{v \in V} \bmu^T_v \mathbf{\Lambda}^{-1}_v (\mathbf{d} - 0.5\bmu_v)
\end{equation}

\subsection{Spreading on multilayer structure}
We use an agent-based version of Susceptible-Infected model \cite{Kermack1927} to simulate the spread across the graph.
According to this model, an agent may be in one of two states, susceptible (S) or infected (I).
At the beginning of the simulation, only one agent is infected -- it is the source.
In the next steps the infected agents interact with their neighbours which as a result may change the susceptible nodes into infected with probability $\beta$ per time step (which is called an infection rate). 
To simplify the tracking of propagation paths and infection times, the dynamics of our model is synchronous, which means that at every time step, all infected nodes try to pass the infection simultaneously to their all susceptible neighbours.

The model is adapted to multi-layer networks, therefore the intra-layer infection rate has a form of vector $\boldsymbol{\beta} = [\beta_1, \beta_2, \dots, \beta_l]$.
We allow replicas to have different state in each layer, which corresponds to a situation when a social media user shares an information on one platform but does not on the other platform.
The strength of coupling between layers is adjusted by the interlayer infection rate $\beta_{inter}$.

\subsection{Evaluation metrics}
Two efficiency measures are used for evaluating the quality of source detection: the average precision and the Credible Set Size at $0.95$ confidence level.
The precision for a single test is defined as the ratio between the number of correctly located sources (i.e., true positives, which here equals either zero or one) and the number of sources found by the method (i.e., true positives plus false positives, which here is at least one).
The tests are repeated multiple times for different origins and many graph realizations (for~synthetic networks) and then the obtained values of precision are averaged.
The Credible Set Size at the confidence level of $\alpha$ ($\alpha\text{-CSS}$) is the size of the smallest set of nodes containing the true source with probability $\alpha$ \cite{Paluch2020}.
In other words this metric describes how many nodes with the highest \textit{score} should be labeled as the source to have the probability $\alpha$ that the true source is among these nodes.
In order to estimate $\alpha\text{-CSS}$ the following procedure is performed.
First, the multiple tests of source detection are conducted and for each test the rank of the true source is computed.
The rank is the position of the node on a list in descending order of the nodes' scores.
Next, $\alpha\text{-CSS}$ is computed as a $\alpha\text{-quantile}$ of the source rank. 

\section{Results}
\begin{figure}[!hbt]
    \centering
    \includegraphics[width=\textwidth]{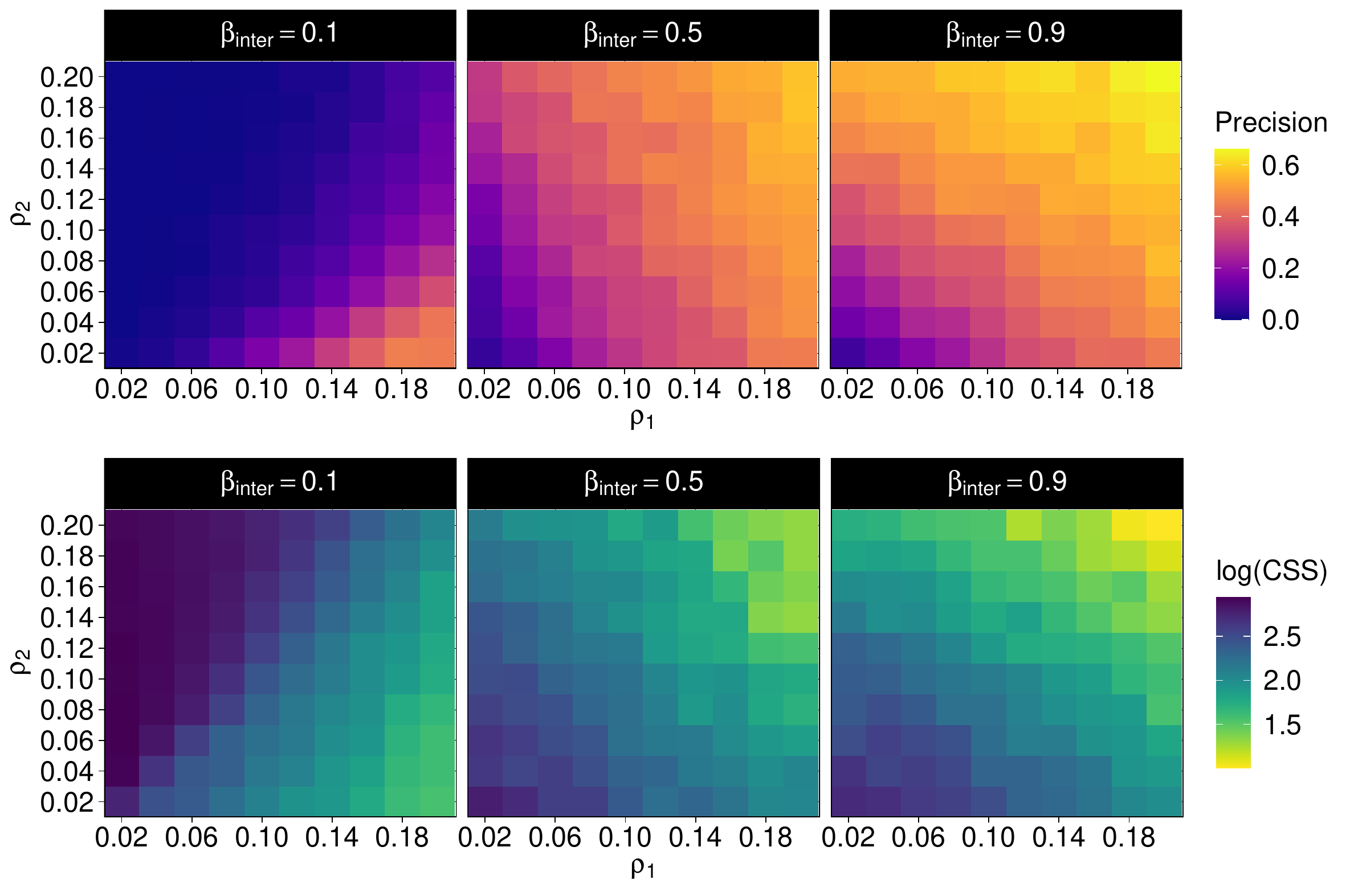}
    \caption{Average precision (higher is better) and 0.95-CSS (lower is better) of source localization in Erd\H{o}s–R\'{e}nyi graph with $L=2$, $n_1=n_2=500$ and $\langle k_1 \rangle = \langle k_2 \rangle = 8$.
    The intralayer spreading rates are $\beta_1 = 0.5$ (source layer) and $\beta_2 = 0.5$.
    We consider three values of interlayer spreading rate $\beta_{inter}$: 0.1 (left), 0.5 (centre) and 0.9 (right).
    The evaluation metrics are computed from $10^3$ realizations.}
    \label{fig:er_density_heatmaps}
\end{figure}
In this section we present the results of two investigations of the impact of multi-layer topology on the quality of the source localization.
The first study focus on the problem of two layers and inspects how the asymmetry between them facilitate or hinder the task of the source detection.
The second study explores the issue of dependency between the number of layers in the network and the quality of the source location.

\begin{figure}[!hbt]
    \centering
    \includegraphics[width=\textwidth]{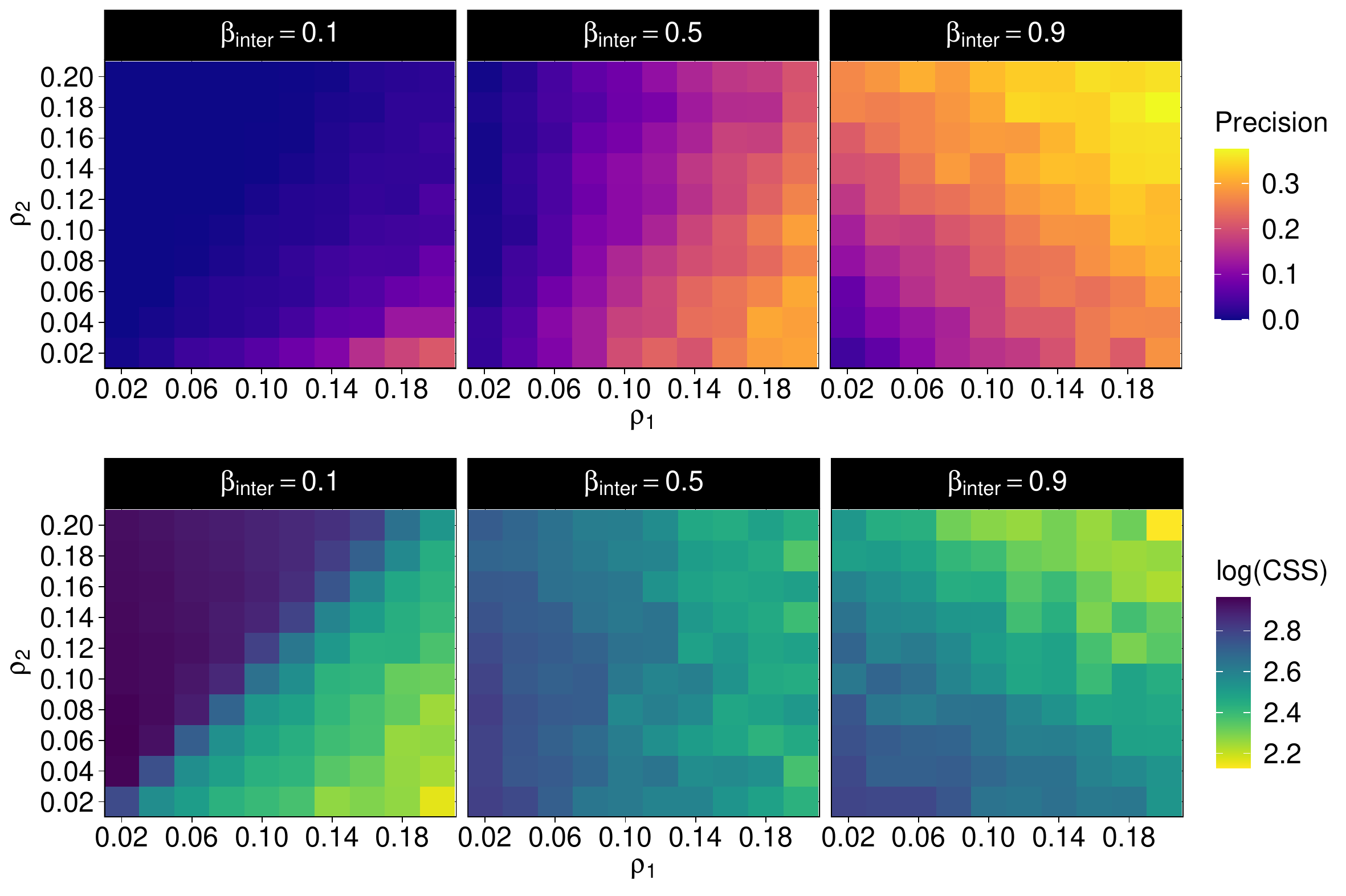}
    \caption{Average precision (higher is better) and 0.95-CSS (lower is better) of source localization in Barab\'{a}si-Albert graph with $L=2$, $n_1=n_2=500$ and $\langle k_1 \rangle = \langle k_2 \rangle = 8$.
    The intralayer spreading rates are $\beta_1 = 0.5$ (source layer) and $\beta_2 = 0.5$.
    We consider three values of interlayer spreading rate $\beta_{inter}$: 0.1 (left), 0.5 (centre) and 0.9 (right).
    The evaluation metrics are computed from $10^3$ realizations.}
    \label{fig:ba_density_heatmaps}
\end{figure}

\subsection{Two layers}
We study the performance of the source localization in bilayer Erd\H{o}s–R\'{e}nyi and Barab\'{a}si-Albert networks for three values of the interlayer infection rate $\beta_{inter}$: low ($0.1$), moderate ($0.5$) and high ($0.9$).
The source of spreading is always placed in Layer 1 which causes that the layers are distinguishable.

In the first experiment we vary the intra-layer infection rates $\beta_1,\beta_2$ from $0.1$ to $0.9$.
The observers are placed randomly with equal density in both layers $\rho_1 = \rho_2 = 0.1$.
As can be seen in Figs. \ref{fig:er_rate_heatmaps} and \ref{fig:ba_rate_heatmaps}, the heat maps of the average precision and 0.95-CSS for weak coupling between layers ($\beta_{inter}=0.1$) differ significantly from the heat maps for moderate and strong couplings.
In the former case, the average precision has two local maximums, one for $\beta_2=0.1$ and second for $\beta_2=0.9$.
For higher values of the inter-layer infection rate $\beta_{inter}$ (centre and right panels in Figs. \ref{fig:er_rate_heatmaps} and \ref{fig:ba_rate_heatmaps}), the characteristics of the  heat maps switches from bimodal to unimodal, with the only one maximum in top right corner ($\beta_1 = \beta2 = 0.9$).
In that case the quality increases overall with the mean value of intra-layer infection rate.

The asymmetry in bi-layer network can be also caused by the difference of densities of observers within the layers, which is shown in Figs. \ref{fig:er_density_heatmaps} and \ref{fig:ba_density_heatmaps}.
Here, the intra-layer spreading rates are equal and moderate $\beta_1 = \beta_2 = 0.5$, but the density of observers $\rho_1,\rho_2$ vary from $0.02$ to $0.2$.
As in the previous experiment, the source of spreading is always placed in Layer 1.
Again, the numerical simulations reveal substantial difference between the networks with weak and strong couplings, but the critical value of $\beta_{inter}$ is lower for Erd\H{o}s–R\'{e}nyi than for Barab\'{a}si-Albert graphs.
In case of weak coupling between layers, the best quality of source location is achieved when the density of observers in the source layer is very high $\rho_1=0.2$, and $\rho_2=0.02$ is very low.
It means, that in this case, the additional observers placed in Layer 2 not only do not help, but also impede the localization of the source in Layer 1.
This behaviour vanishes when $\beta_{inter} > \beta_{critical}$, because then, the quality of source detection increase with the density of observers in any layer.

\begin{figure}[!hbt]
    \centering
    \begin{subfigure}[b]{0.48\textwidth}
        \caption{Erd\H{o}s–R\'{e}nyi}
        \includegraphics[width=\textwidth]{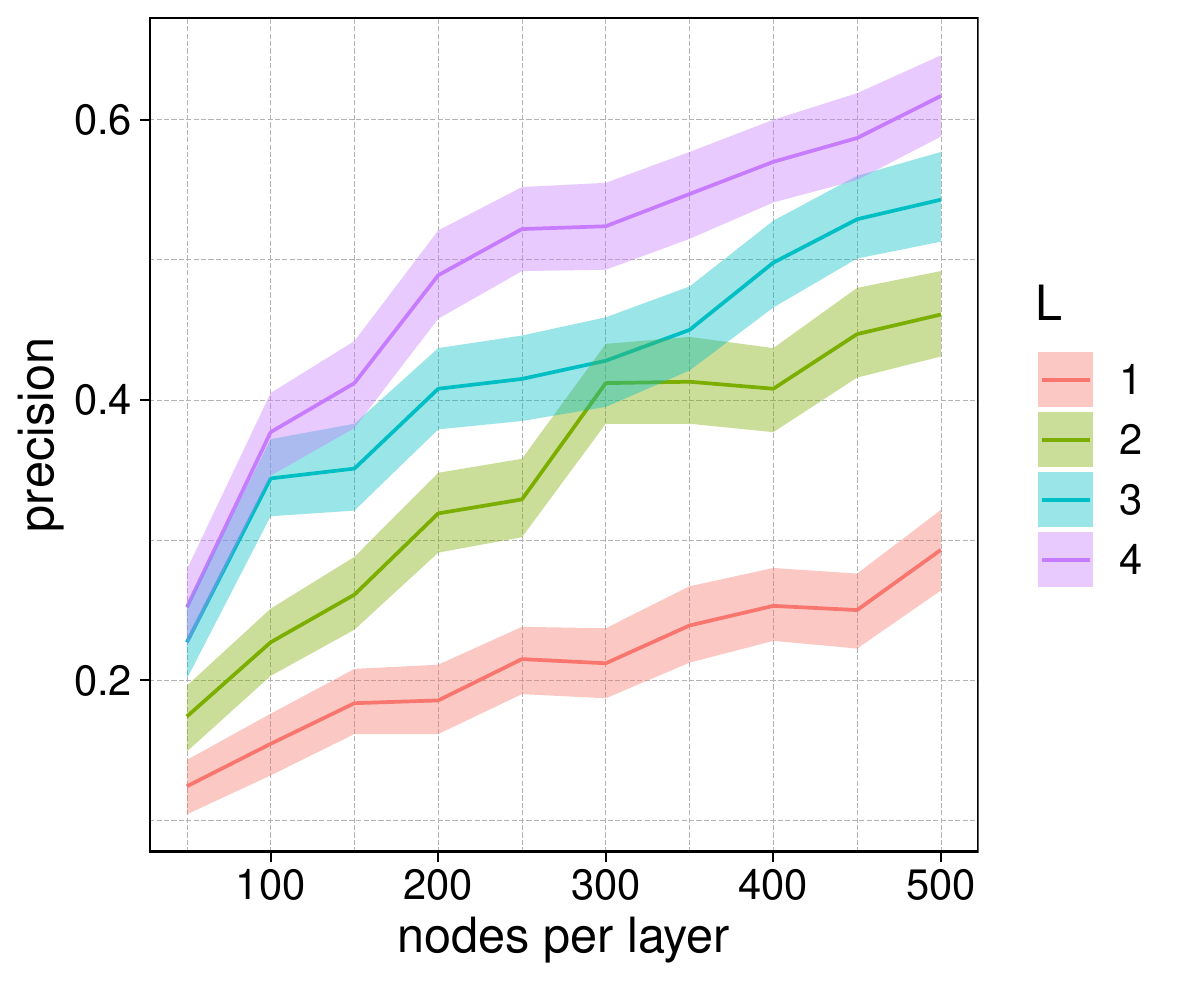}
    \end{subfigure}
     ~
    \begin{subfigure}[b]{0.48\textwidth}
        \caption{Barab\'{a}si-Albert}
        \includegraphics[width=\textwidth]{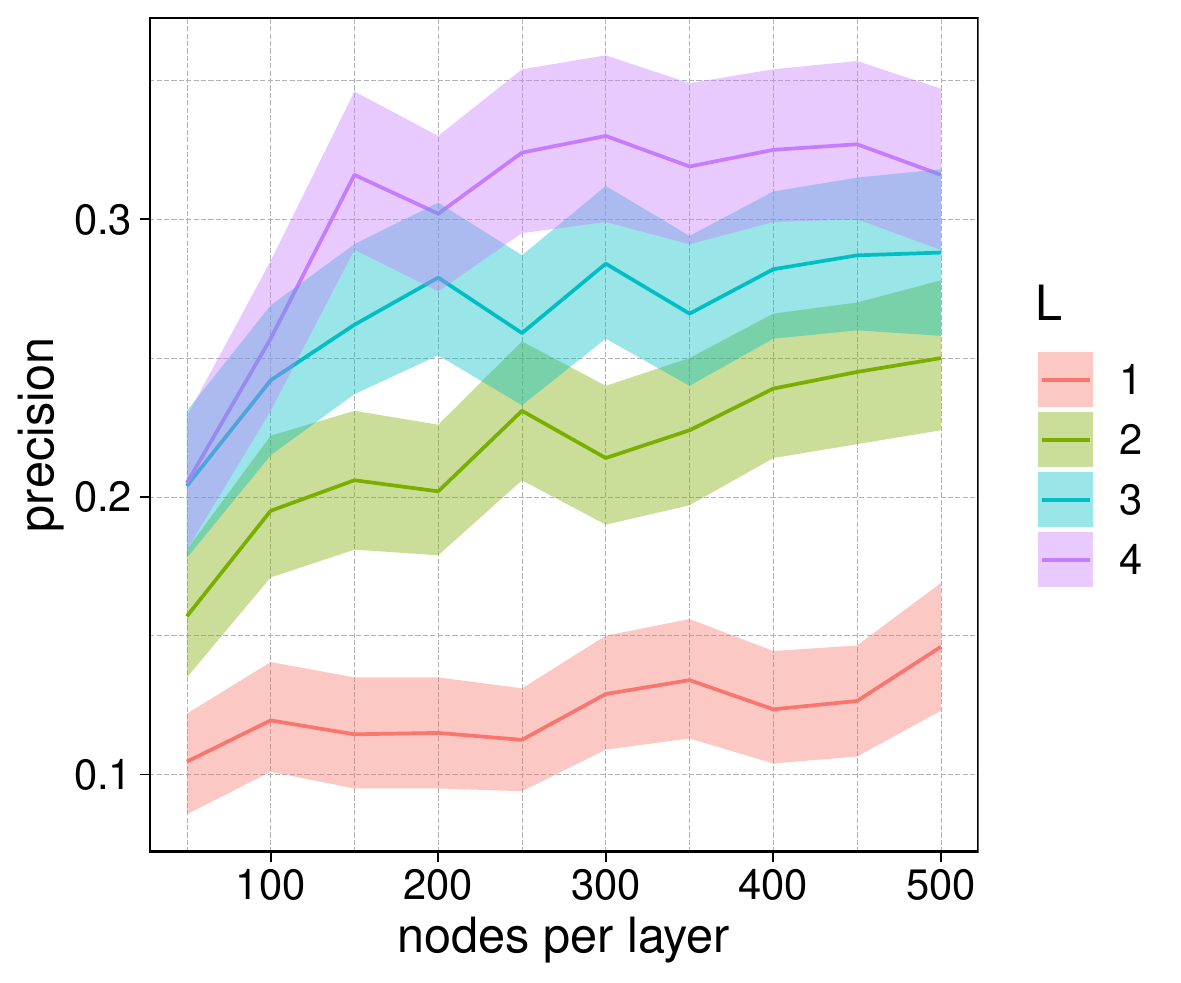}
    \end{subfigure}
    ~
    \begin{subfigure}[b]{0.48\textwidth}
        \includegraphics[width=\textwidth]{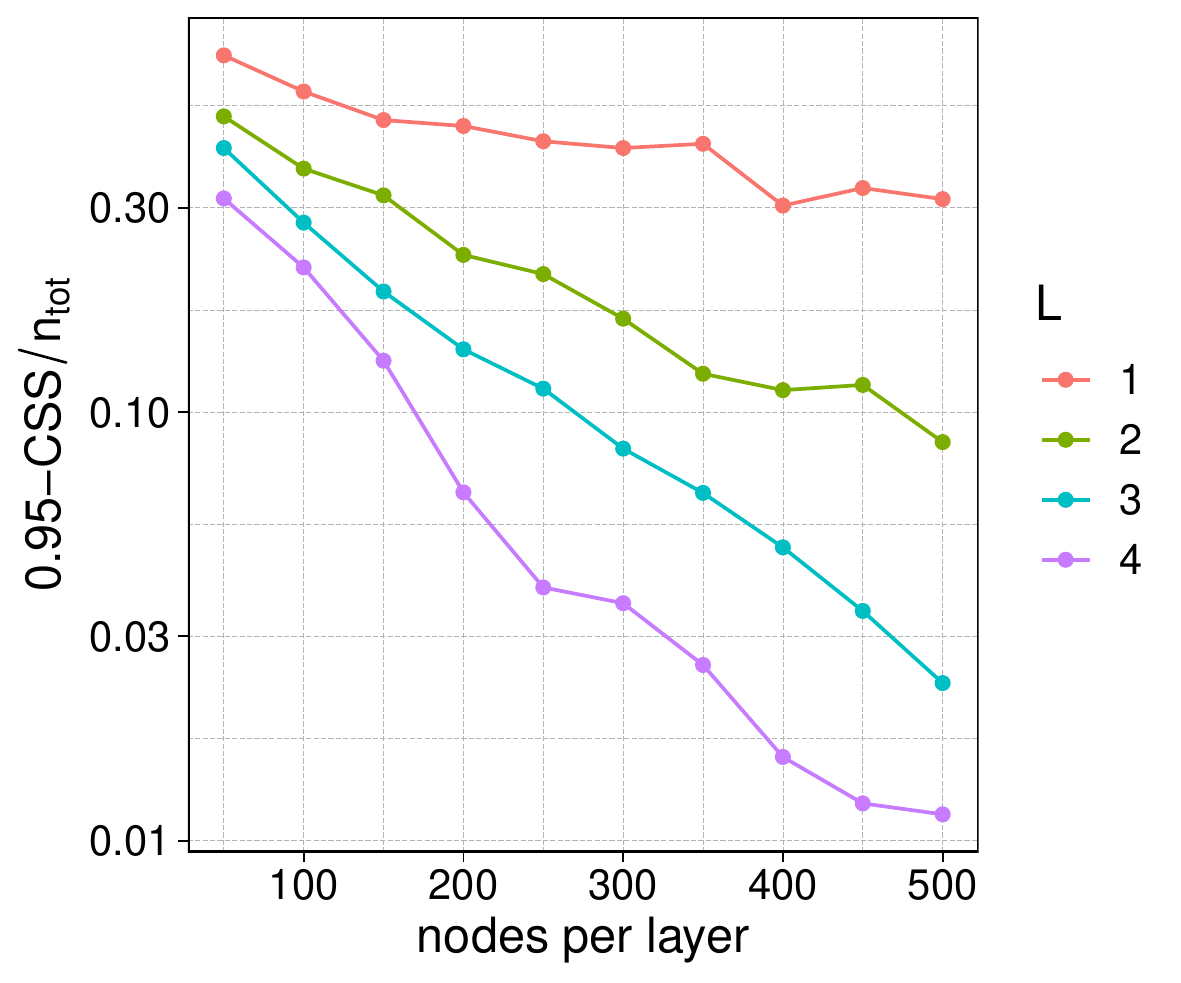}
    \end{subfigure}
    ~
    \begin{subfigure}[b]{0.48\textwidth}
        \includegraphics[width=\textwidth]{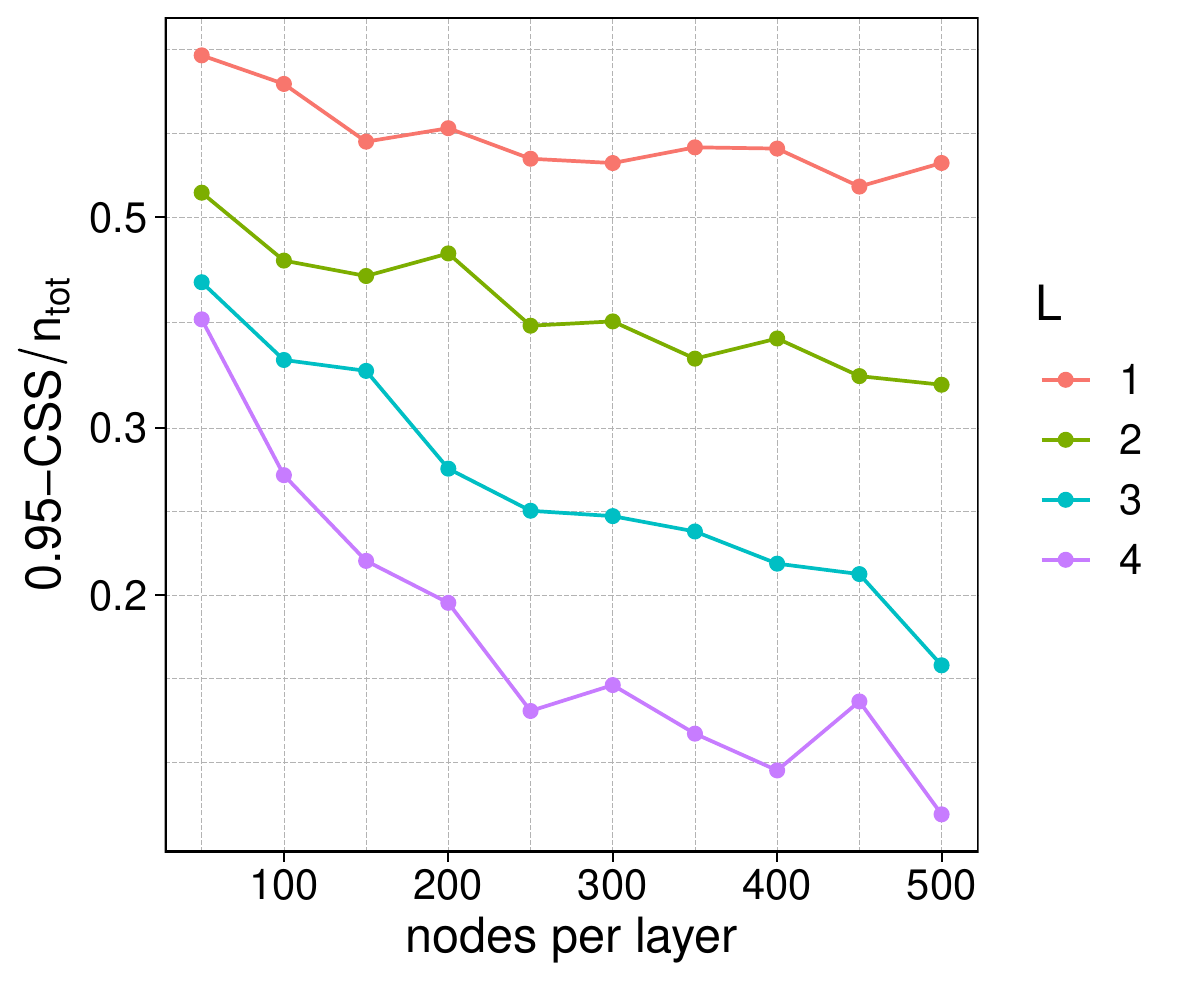}
    \end{subfigure}
    \caption{The quality of source localization in Erd\H{o}s–R\'{e}nyi graph (left) and  Barab\'{a}si-Albert model (right) with different number of layers.
    The number of nodes per layer is the same for all networks, regardless of the number of layers $L$.
    The average degree $\langle k \rangle = 8$, density of observers observers $\rho=0.1$ and intralayer infection rate $\beta_{intra}=0.5$ are the same in each layer.
    The interlayer infection rate is $\beta_{inter}=0.8$.}
    \label{fig:constant_nodes_per_layer}
\end{figure}

\subsection{More than two layers}
\begin{figure}[!hbt]
    \centering
    \begin{subfigure}[b]{0.48\textwidth}
        \caption{Erd\H{o}s–R\'{e}nyi}
        \includegraphics[width=\textwidth]{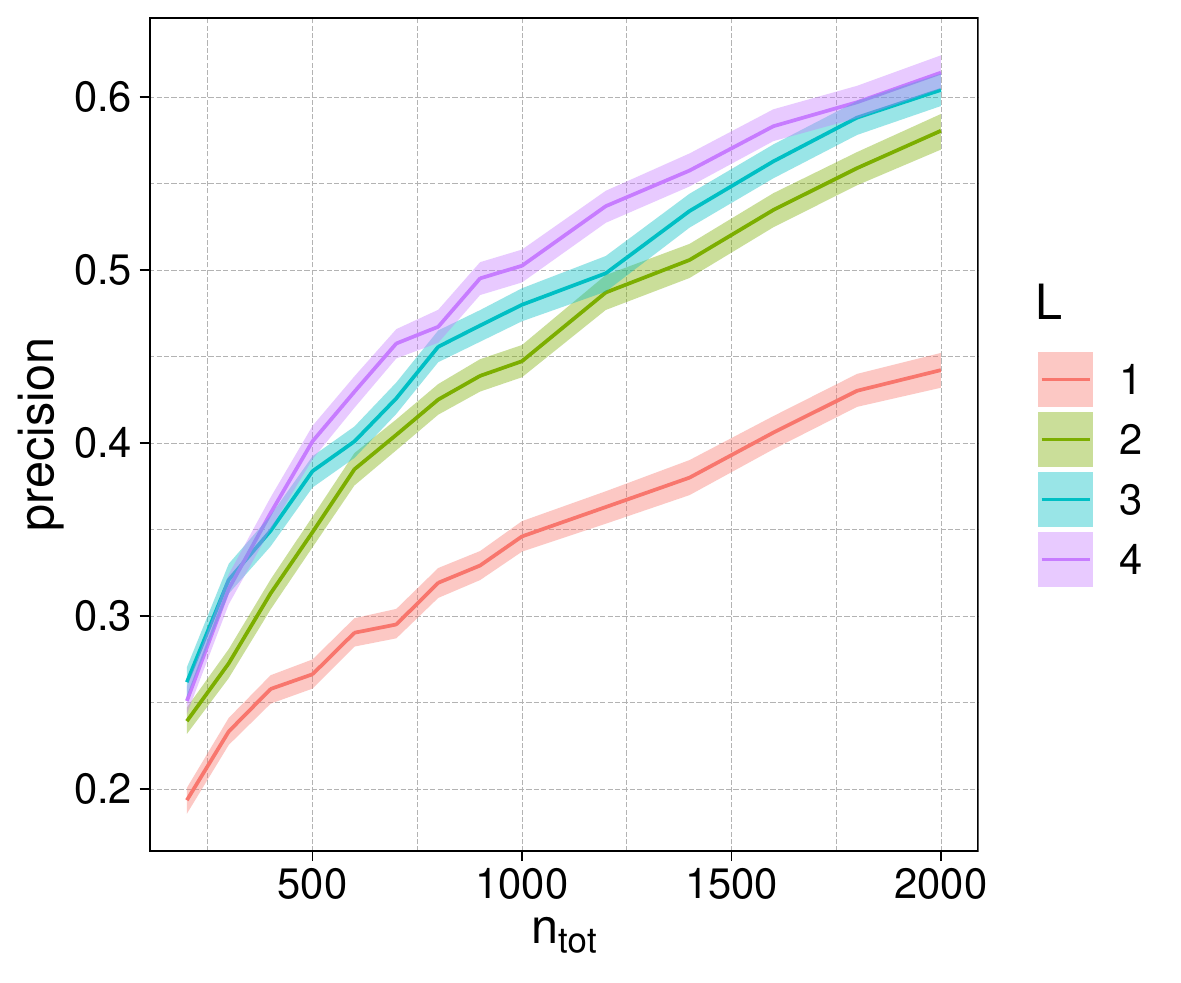}
    \end{subfigure}
        ~
     \begin{subfigure}[b]{0.48\textwidth}
        \caption{Barab\'{a}si-Albert}
        \includegraphics[width=\textwidth]{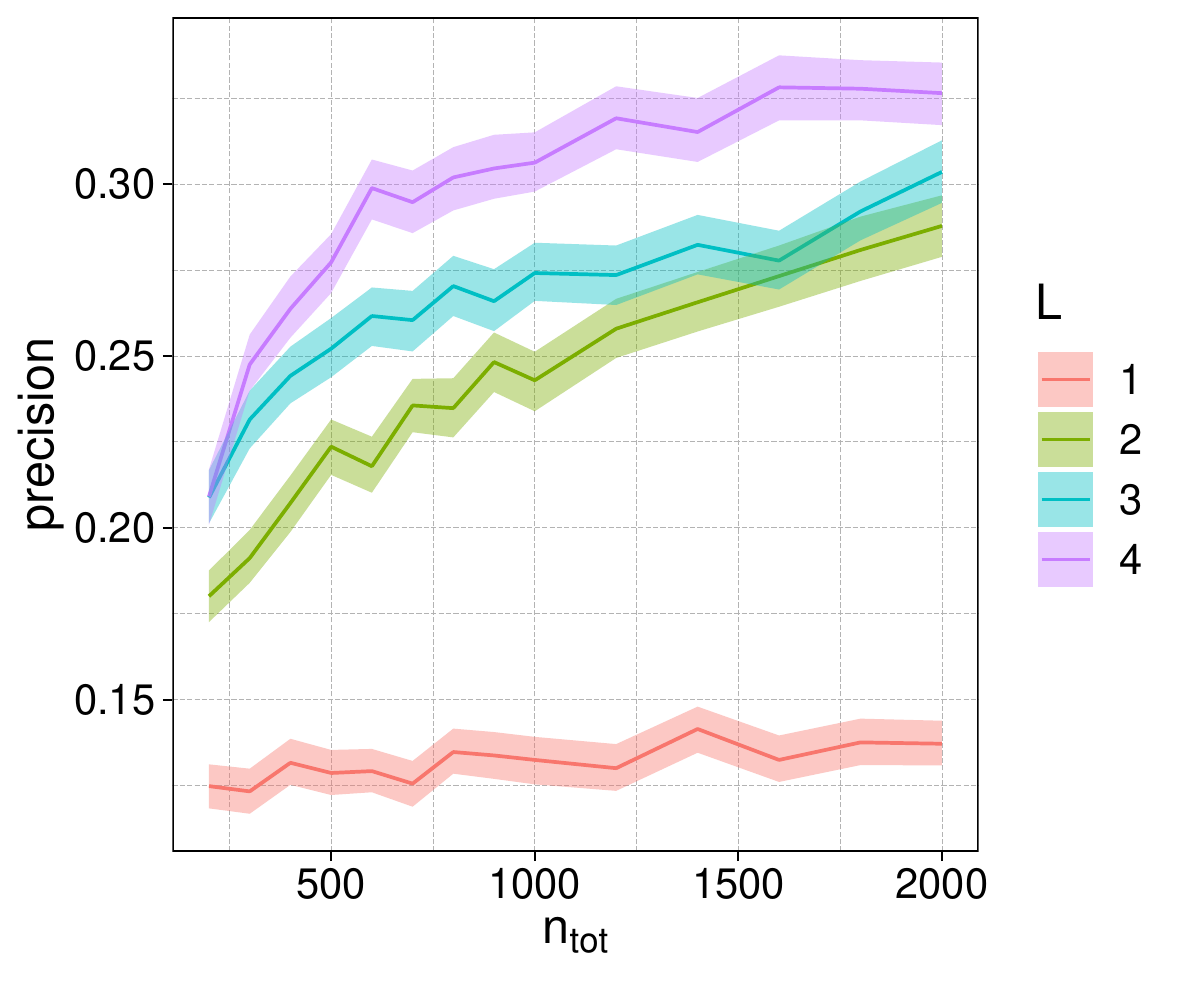}
    \end{subfigure}
    ~
    \begin{subfigure}[b]{0.48\textwidth}
        \includegraphics[width=\textwidth]{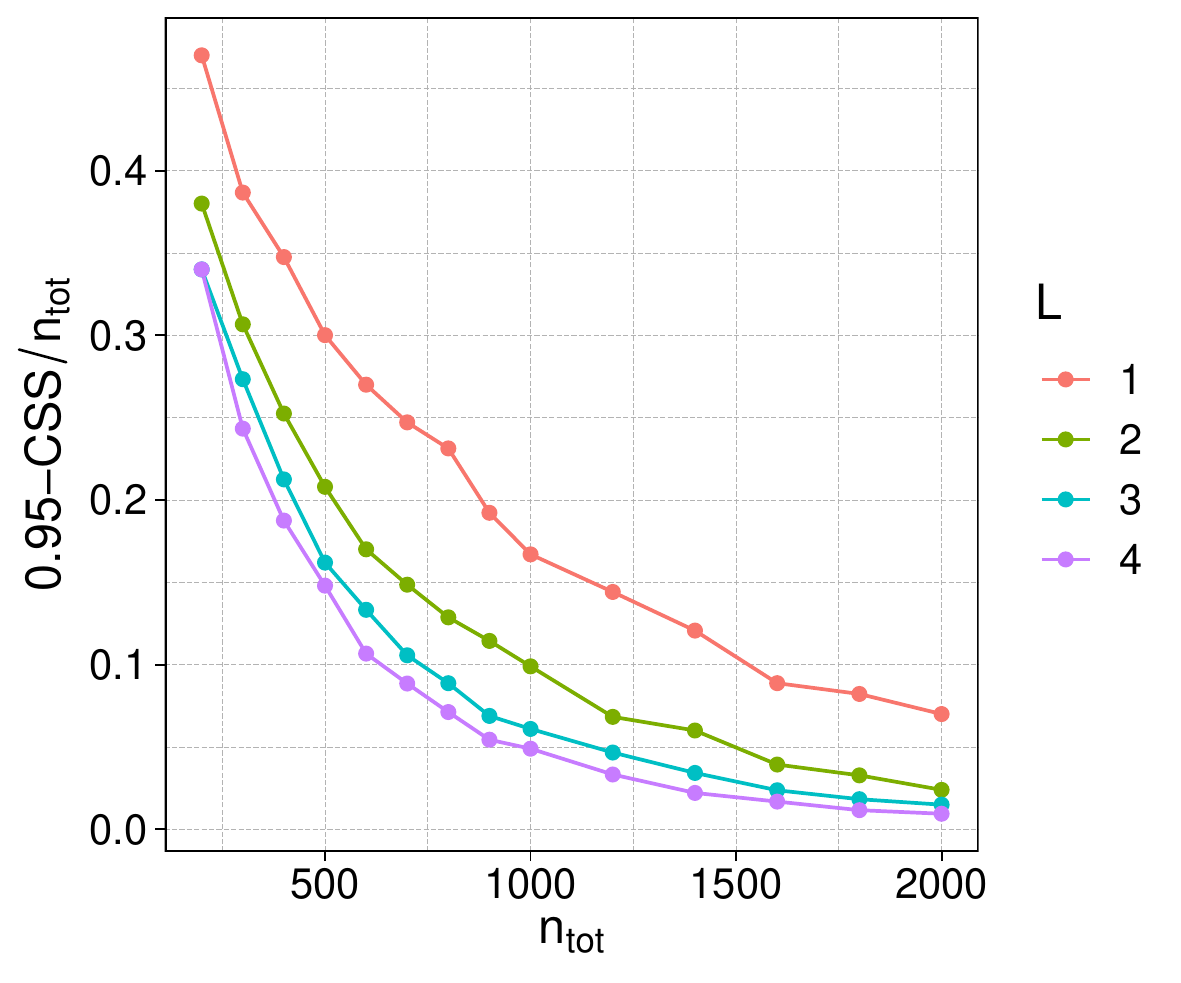}
    \end{subfigure}
    ~
    \begin{subfigure}[b]{0.48\textwidth}
        \includegraphics[width=\textwidth]{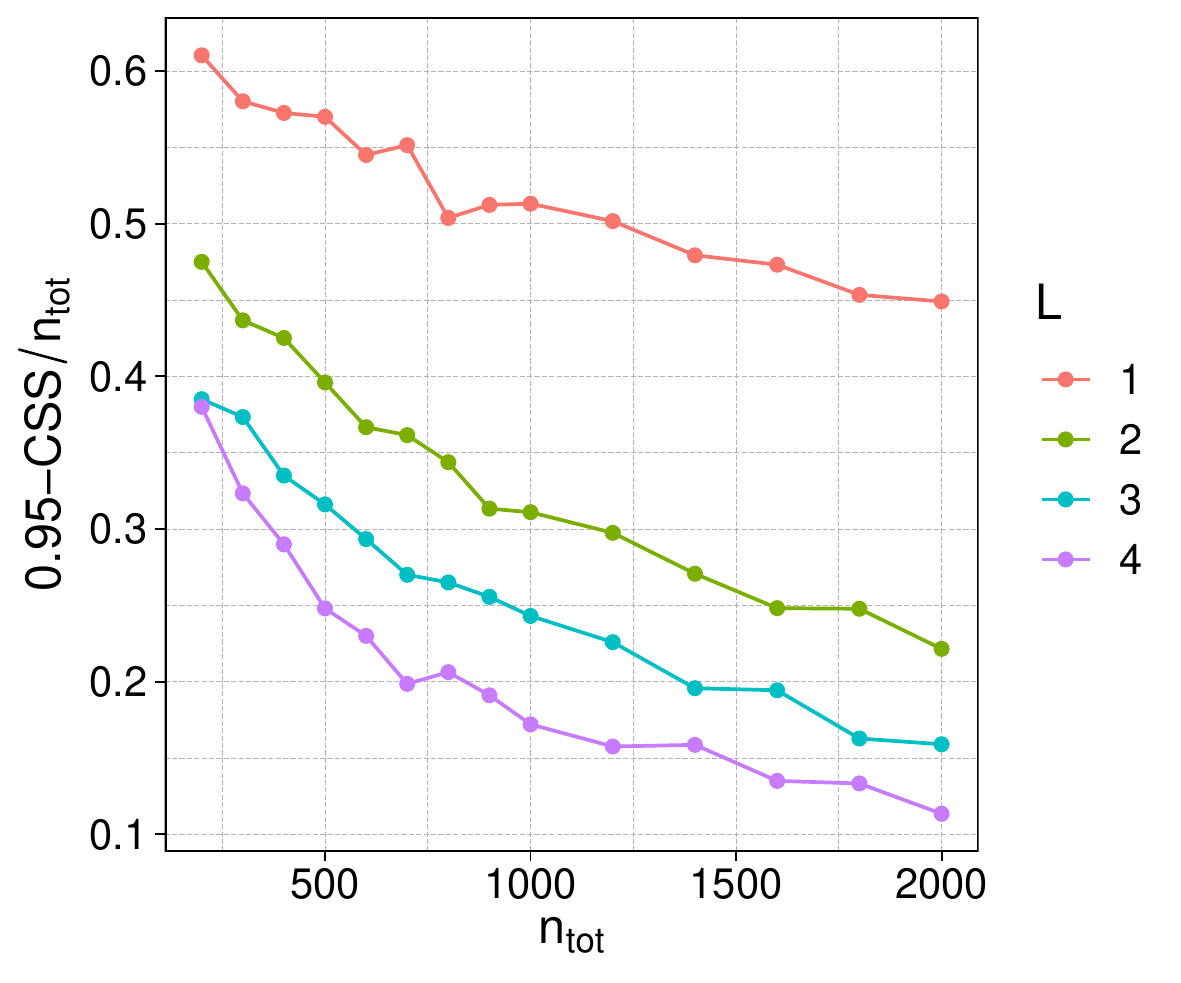}
    \end{subfigure}
    \caption{The quality of source localization in Erd\H{o}s–R\'{e}nyi graph (left) and  Barab\'{a}si-Albert model (right) with different number of layers.
    The number of nodes per layer is inversely proportional to the number of layers $L$.
    The average degree $\langle k \rangle = 8$, density of observers observers $\rho=0.1$ and intralayer infection rate $\beta_{intra}=0.5$ are the same in each layer.
    The interlayer infection rate is $\beta_{inter}=0.8$.}
    \label{fig:constant_total_size}
\end{figure}

We compare the performance of the source location algorithm for the networks with different number of layers according to two schemes.
In the first one, all considered systems have the same number of nodes per layer $n_l$ but they differ in total number of nodes $n_{tot}$, e.g. a four-layer graph has twice as many nodes as a two-layer one.
In the second scheme, the total number of nodes $n_{tot}$ is the same for all compared networks, which means that a two-layer graph has twice as many nodes per layer as a four-layer one.
The results presented in Figs. \ref{fig:constant_nodes_per_layer} and \ref{fig:constant_total_size} show strong influence of the number of layers on the average precision and 0.95-CSS for both the case of fixed $n_l$ and fixed $n_{tot}$.
A particularly large increase in the performance of the source location is observed after changing from one layer to two-layer graph.
One can speculate that in the first scheme, the one with fixed number of nodes per layer $n_l$, the networks with larger number of layers have also more observers, but this is not the case in second scenario, when the number of observers depends only on $n_{tot}$, which on the other hand is independent of the number of layers $L$.

\section{Discussion} 
Nowadays it is not uncommon to have to deal with dissemination on multi-layer networks of many kinds and often finding the source of said propagation can be a crucial task. Whether it is a biological pandemic or a virtual infodemic it is important to have tools allowing for locating the origin. This task can be very challenging and especially on multi-layer systems. In this paper we tackle this exact problem with a maximum likelihood approach that we extend to be operational on multi-layer graphs. We test our method for source location estimation on synthetic networks outlining  its potential strengths and limitations. We also observe some non-trivial and perhaps surprising phenomena where the more of the system one observes the worse the results become.

We first conduct several tests on synthetic networks - Erd\H{o}s–R\'{e}nyi and Barab\'{a}si-Albert - starting in bi-layer scenarios and later moving onto many-layer examples.

In the studies cases of two layer systems on one hand, we observe that the weak coupling between layers can act as an information buffer or a dam - if the $\beta_{inter}$ is low then (almost) no matter how many observers we have in the second (non-source) layer the quality of localization does not improve, i.e. these observers do not provide meaningful information. There is enough of a \textit{separation} of these observers from the source that they are utterly hopeless in their task.

On the other hand, this separation can become dramatic enough to \textit{misinform} the observers. We observe this when infection rates are sufficiently large in both layers and $\beta_{inter} < \beta_{critical}$. For such a case this buffering behaviour of a sufficiently weak coupling retards our localization capabilities - our observers have a distorted view of the propagation and the more of them we have the worse our precision.

It is worth noting that these effects are to some extent a consequence of the propagation dynamics we chose. In Susceptible-Infected model the information traversal times have a geometric distribution contingent on the $\beta$ parameter. As such the longer the route (topological length of the traversal path) and lower the $\beta$ the higher the variance of the arrival time.

We also discover that in the $(\beta_1, \beta_2)$ phase space the transition in regard to $\beta_{inter}$ not only shifts the precision maximum from high to low values of $\beta_2$ as $\beta_{inter}$ decreases but also splits it into two. That is for small values of the coupling parameter we observe two - not one as in other cases - high precision score ``islands'': one for  high $\beta_2$ and $\beta_1$ (the typical one) and the other for low values of $\beta_2$ and high values of $\beta_1$. We interpret the first local maximum (both $\beta$ parameters are high) as simply a counter-balancing of the buffering effect described earlier via the weak coupling. The nature of the second maximum for low values of $\beta_2$ is less clear. Perhaps simply the observers in the source layer have sufficiently high quality of information that the second layer's observer cannot impede on the results.

As we increase the number of layers we shift our focus onto studying the precision in terms of the number of said layers and the system size, the latter in two forms - fixed total number of nodes and fixed number of nodes per layer.

In tested scenarios of high inter-layer coupling parameter in all cases we observe an increase in precision as the system size increases and as the number of layers increases. This can perhaps be counter-intuitive as one would expect that the added complexity of multi-layer systems would impede on our localization capabilities especially considering our previous analysis of bi-layer scenarios.

In the first case we believe that this can be explained as we simply have more observers. Since we fix the \textit{density} of observers it is similar to what we had observed in some of our previous work \cite{gradient-pinto}. Namely, even for a single layer case as the system size increases so does the localization performance when the observer density is kept fixed. This is due to the fact that in order to keep a stable precision score with increasing network size the observer set size does not grow linearly which in turn is a consequence of average shortest path lengths in a small world graph growing logarithmically with graph size.

In the second case we suspect that this is an effect also similar to one we had seen before - when the precision increases with the mean degree until some $\langle K \rangle_{optimal}$ and then decreases. In the multi-layer scenario we believe that as we add more layers we also add more available nodes close to the source that can become observers. Namely, the second layer adds paths of length $2$, the third layer of length $3$ and so on which is very similar to simply increases the mean degree.

We are aware that the results presented here are far from exhaustive. Source localization on multi-layer networks is both conceptually and computationally intensive task and many questions still remain as they fall beyond the scope of this paper. In the future we plan on studying this topic further and in particular to look more closely on how the coupling parameter in many layer scenarios impacts the performance of the maximum likelihood estimation and whether we can observe the conjectured effect of eventual decreasing of this performance as the number of layers becomes sufficiently large. Nevertheless, we find this to be an important subject of study in the field of complex networks and believe our results will prove useful to the scientific community and beyond.

\section*{Acknowledgments}
The work was partially supported by the National Science Centre, Poland Grant No. 2015\-/19\-/B\-/ST6\-/02612.
R.P. was partially supported by the National Science Centre, Poland, agreement No 2019/\-32/\-T/\-ST6/\-00173, and by PLGrid Infrastructure.
J.A.H. was partially supported by the Russian Science Foundation, Agreement No 17-71-30029 with co-financing of Bank Saint Petersburg, Russia.

\typeout{}
\bibliographystyle{ieeetr}
\bibliography{references}

\end{document}